\documentclass{article}
\usepackage{amsmath, amssymb, enumerate, amsthm}
\usepackage{amsfonts, times, graphicx, float}
\usepackage{caption}

\usepackage{amsmath,amsfonts,amssymb,amsthm,booktabs,color,epsfig,graphicx,hyperref,url}

\usepackage{graphicx,subfigure,color,enumerate}
\usepackage[utf8]{inputenc}
\usepackage{mathptmx}
\usepackage[11pt]{moresize}
\newtheorem{definition}{Definition}

\newtheorem{example}{Example}

\begin{document}

\title{Sampling the Swadesh List to Identify Similar Languages with Tree Spaces}
\author{Garett Ordway\\Florida State University, Florida, U.S.A.
\and Vic Patrangenaru\\Florida State University, Florida, U.S.A.}
\date{April 9, 2024}
\maketitle

\begin{abstract}
Communication plays a vital role in human interaction. Studying language is a worthwhile task and more recently has become quantitative in nature with developments of fields like quantitative comparative linguistics and lexicostatistics. With respect to the authors own native languages, the ancestry of the English language and the Latin alphabet are of the primary interest. The Indo-European Tree traces many modern languages back to the Proto-Indo-European root. Swadesh's cognates played a large role in developing that historical perspective where some of the primary branches are Germanic, Celtic, Italic, and Balto-Slavic. This paper will use data analysis on open books where the simplest singular space is the 3-spider - a union $\mathcal{T}_3$ of three rays with their endpoints glued at a point 0 -  which can represent these tree spaces for language clustering. These trees are built using a single linkage method for clustering based on distances between samples from languages which use the Latin Script. Taking three languages at a time, the barycenter is determined. Some initial results have found both non-sticky and sticky sample means. If the mean exhibits non-sticky properties, then one language may come from a different ancestor than the other two. If the mean is considered sticky, then the languages may share a common ancestor or all languages may have different ancestry.
\end{abstract}


%\end{frontmatter}

\section{Introduction\label{s:1}}
Whether you believe the story that God confused the languages of mankind at the Tower of Babel, that the multitude of modern-day languages evolved from one common language as migration transpired (monogenesis), or that distinct languages have existed since the beginning of the human species (polygenesis), language is ubiquitous throughout history. Woodard summarizes many scholars with his introduction in \textit{The Ancient Languages of Europe} \cite{Woodard:2010}. Several languages have records beginning more than 4500 years ago (c. 2600 BC). Algeo in \textit{The Origins and Development of the English Language} argues that human language is ``fundamentally different" from animal communication. While apes have shown the most progress of animals in learning communication, ``it is a far cry from the fullness of a human language." Human language is special\cite{AlgeoPyles:2010}. \\
As different people groups began to interact, translation of languages certainly began quickly. \textit{The Sumerian Epic of Gilgamesh} is one of the most commonly known early literary works that has been found in multiple languages. A more modern and certainly more famous ancient translation is the Rosetta Stone believed to be from 196 BC with an Ancient Greek and two Ancient Egyptian languages. Augustine of Hippo in AD c.420 wrote in \textit{City of God} about the Tower of Babel and discussed the many languages that needed translations of the Bible \cite{AugustinevonBabcockColemanAugustine:2021}. Campbell's \textit{Historical Linguistics: An Introduction} connects philology to linguistics which encapsulates the modern-day study of language. Fields such as quantitative comparative linguistics, historical linguistics, glottochronology, and lexicostatistics have searched for insights with varying methods. While Morris Swadesh was not the first linguist (the pioneer of monogenesis was Alfredo Trombetti \cite{Trombetti:1962} in 1905), he has made perhaps the biggest contribution. Swadesh's initial study was of America's indigenous languages. His most famous contributions began in the 1950s with the publication of the first Swadesh List \cite{Swadesh:1952}. Subsequent Swadesh lists contained more or fewer or different cognate words. Crystal's \textit{A Dictionary of Linguistics and Phonetics} defines a cognate as ``a language or linguistic form which is historically derived from the same source as another language/form"\cite{Crystal:2024}. Modern languages identify which ancient language they received words from or which other modern languages they are similar to using these cognates. In English, some of these cognate words from Swadesh's 207 list are ``where," ``big," ``child," and ``to think." From a monogenesis perspective, the search continues for a connection of all world languages to a ``Proto-Human" language in a famous work from Merritt Ruhlen \cite{Ruhlen:1994}. As computational power increases and archaeology continues uncovering ancient civilizations, hope grows. \\
In the search for the original language, scholars identify several Proto-languages such as Proto-Indo-European and Proto-Afroasiatic. While these most ancient languages are only hypothesized, they do allow for adding structure to the study of language. The Indo-European language family's prominence in Western Civilization and preserved writings for millennia has produced a fairly settled phylogenetic tree with several main branches being Germanic, Latin (Italic), Hellenic (Greek), Celtic, Balto-Slavic, and Indo-Iranian. The phylogenetic tree should depict language origins and groupings of similar languages. Trees can be developed through a variety of methods ranging from completely subjectively to expert opinion to data modeling.

These trees have their own roots from scientists' studies of natural world in the 1800s trying to trace the lineage of plants and animals \cite{Darwin_2024} \cite{Archibald_2008}. A phylogenetic tree attempts to trace the evolutionary ancestry of plant and animal species. The species is the final branch of the tree, and branches connect at nodes. In the animal kingdom, a taxonomic rank is imposed to distinguish the different species within a genus from other genera and different genera within a family from other families and so on \cite{Mishler_Wilkins_2018}. While taxonomy provides organization for the animal and plant kingdoms, phylogenetics attempts to provide answers for the evolutionary ancestry. How the phylogenetic tree is built depends on the method of analysis used. Early trees (mid-19th century) were built using only similar characteristics, but as the field of phylogenetics developed, trees grew in complexity using techniques from statistics such as Maximum Likelihood \cite{Edwards_1963} \cite{NEYMAN_1971}, Nearest Neighbor Interchange \cite{ROBINSON_1971}, and Bayesian Inference \cite{Li_2000}\cite{Mau_2004}\cite{Rannala_Yang_1996}. Phylogenetic trees have thus become very well-known even to laymen through their development in systematic biology and use in science textbooks and, more recently, investigations related to the SARS-Cov-2 virus outbreak \cite{Li_2020}\cite{Chen_2023covid}\cite{Attwood_2022} \cite{Zhao_2022}. A SARS-Cov-2 phylogenetic tree can be made since the approximately 30,000 letter sequence is widely studied illustrating the rapid effects of microevolution.

Phylogenetic trees are being extended to include work on branch length and phylogenetic networks. Phylogenetic trees must then have topological properties and can be studied from that perspective. Billera, Holmes, and Vogtmann's critical work ``Geometry of the Space of Phylogenetic Trees" gave significant mathematical insight for future studies \cite{BILLERA:2001}. They gave the idea that the usual Euclidean distance can be used to calculate distances for points in the same orthant while adding two straight segments for points in different orthants can give a distance as well. Additionally, they showed that a centroid can give a consensus of set of trees. Later, Hotz et al extended this work to the open book and proved the Sticky Central Limit Theorem \cite{Hotzetal:2013}. Paradis and Binet et al both discuss the idea of branch length carrying some importance on the phylogenetic tree \cite{PARADIS2016} \cite{Binet_2016}. For evolutionary theory, a longer branch should indicate more genetic distance. Huson and Bryant consider phylogenetic trees to be one type of phylogenetic network extending the idea to allow branches to interact with one another separately from connecting to the node \cite{Huson_2005}. Phylogenetics continues to develop with key contributions over the last several decades.

But biology is not the only possible application of phylogenetics. Human language has also undergone significant change over the past millennia. While Swadesh was one of the first linguists to popularize the language tree to show modern languages connection to ancient languages, some work had been done previously (prior to 1950), and much work has been done since in applying a wide variety of techniques. Early work by Kroeber calculated similarity coefficients for languages based on a plus or minus approach denoting whether a characteristic was present or absent and counting some number of times the characteristic was present in both, present in one, or absent in both \cite{Kroeber_1937}. Swadesh's cognates stirred much discussion from the 1950s onward prompting discussion of choice of native words and synonyms and gave birth to the lexicostatistical approach of which Dyen's 1965 work is paramount \cite{dyen1965lexicostatistical}. Dyen published with Kruskal and Black in 1992 a thorough analysis of the IndoEuropean tree including the new \emph{box diagram} \cite{Dyen_1992}. They used hierarchical clustering with lexicostatistical percentages calculated from similar or different cognate words. Glottochronology attempts to determine dates of linguistic evolution through rates of change. Swadesh, Lees \cite{Lees_1953}, and Hymes \cite{Hymes_1973} all provided early contributions in the field with Chretien emphasizing the mathematics of the function \cite{chretien_1962}. Modern studies by Gray and Atkinson attempt to go farther back in history to original emergence of the Indo-European languages \cite{Gray_Atkinson_2003}. Other approaches to language evolution involve concern with borrowed words due to cultural interactions and phonetic changes over time, among others, necessitating a variety of methods similar to those mentioned for biological phylogenetics earlier.

Some examples of phylogenetic trees for language are presented next. Algeo gives the following tree for the Indo-European family based on comparative linguistics \cite{AlgeoPyles:2010}.

\begin{figure}[H]
\begin{center}
\includegraphics[scale = 0.45]{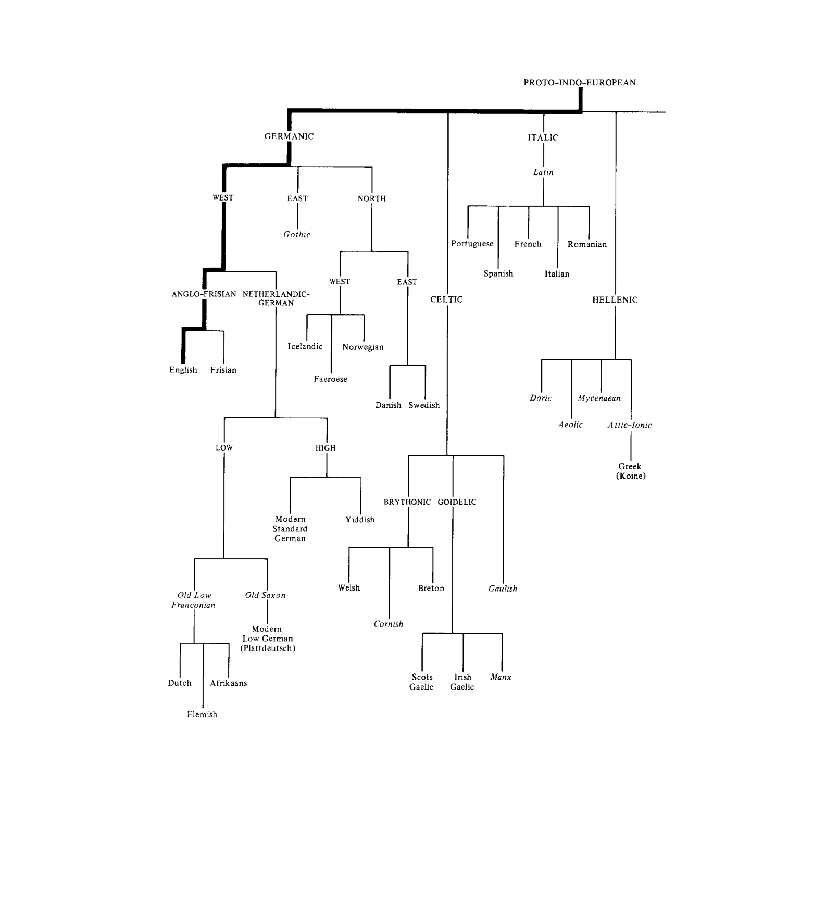}
\end{center}
\caption{Indo-European Languages Tree-part I}
\label{fig:Lang_tree_1}
\end{figure}

\begin{figure}[H]
\begin{center}
\includegraphics[scale = 0.45]{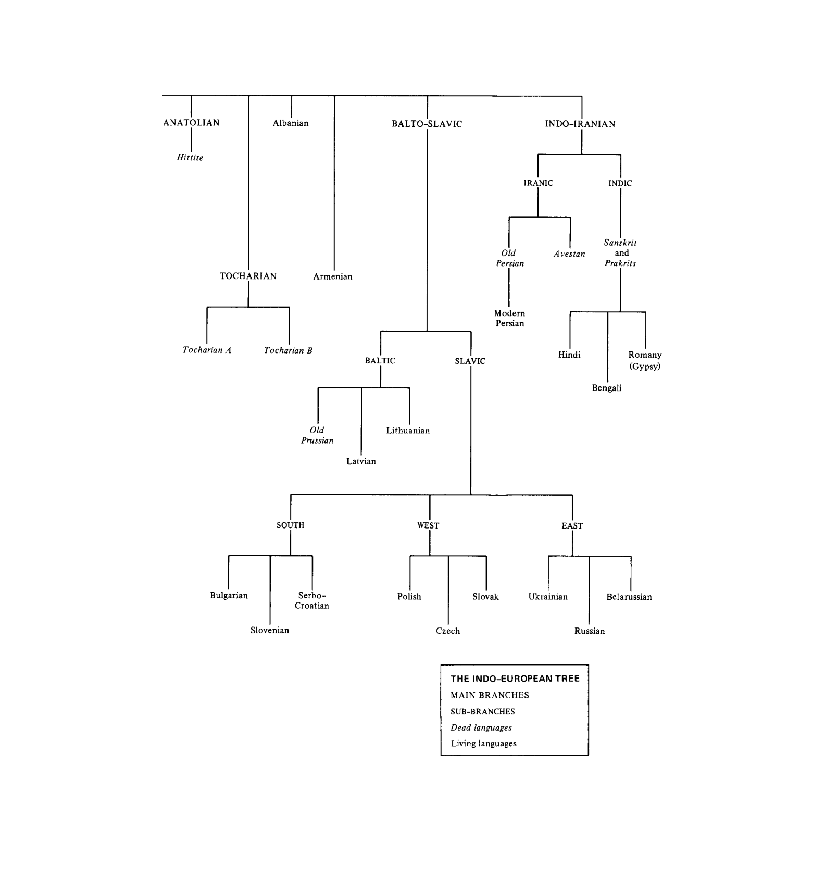}
\end{center}
\caption{Indo-European Languages Tree-part II}
\label{fig:Lang_tree_2}
\end{figure}

Johnson and Wichern's use of the first letter of the ten numerals gives this tree diagram \cite{JohnsonWichern:2019}.
\begin{figure}[H]
\begin{center}
\includegraphics[scale = .6]{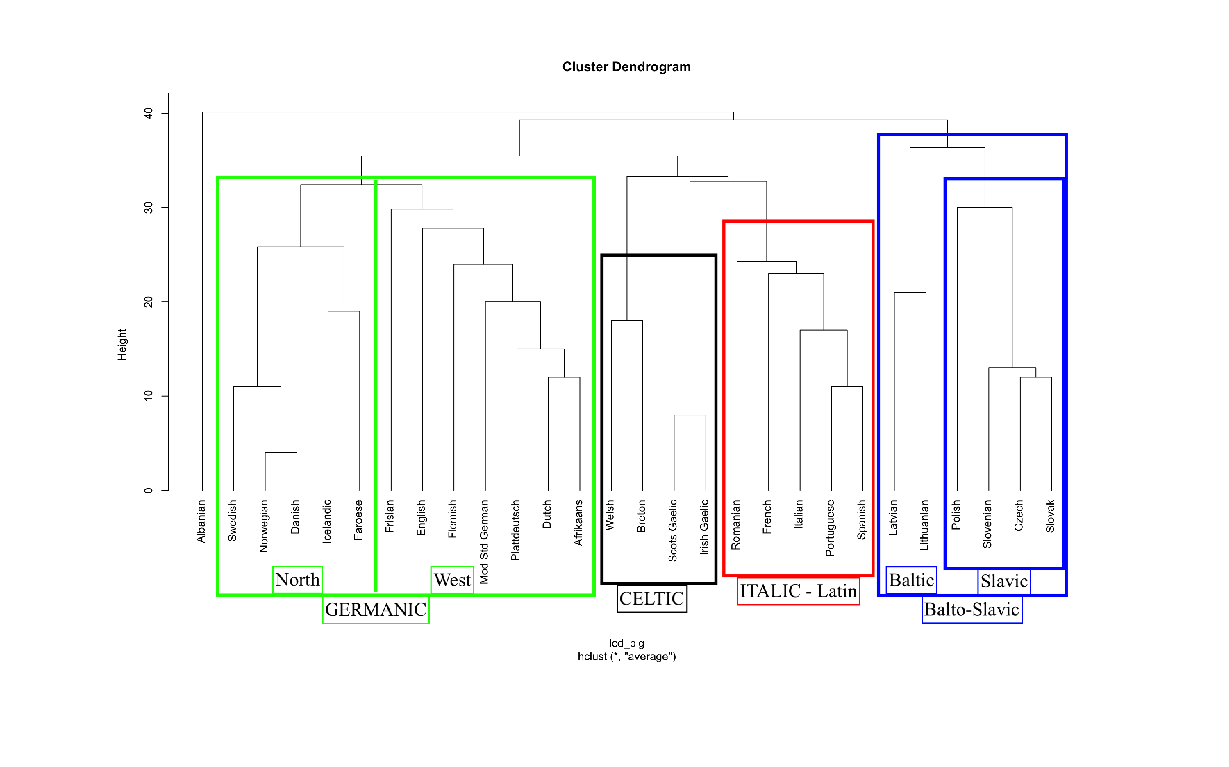}
\end{center}
\caption{Partial Indo-European Languages Tree from numbers}
\label{fig:Lang_tree_3}
\end{figure}

A tree for Indo-European languages, due to an anonymous author, is displayed below can be found at 
\cite{An:2024}
\begin{figure}[H]
\begin{center}
\includegraphics[scale = .8]{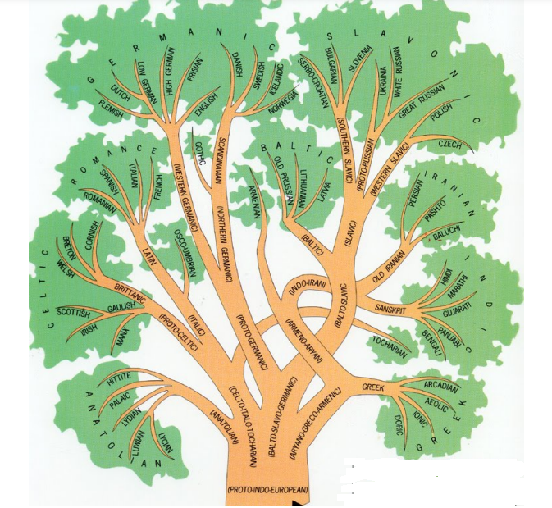}
\end{center}
\caption{Indo-European Languages Tree Tree}
\label{fig:Lang_tree_4}
\end{figure}

\begin{figure}[H]\label{f:tree-spaces}
%$T_3$.
\includegraphics[scale = 0.20]{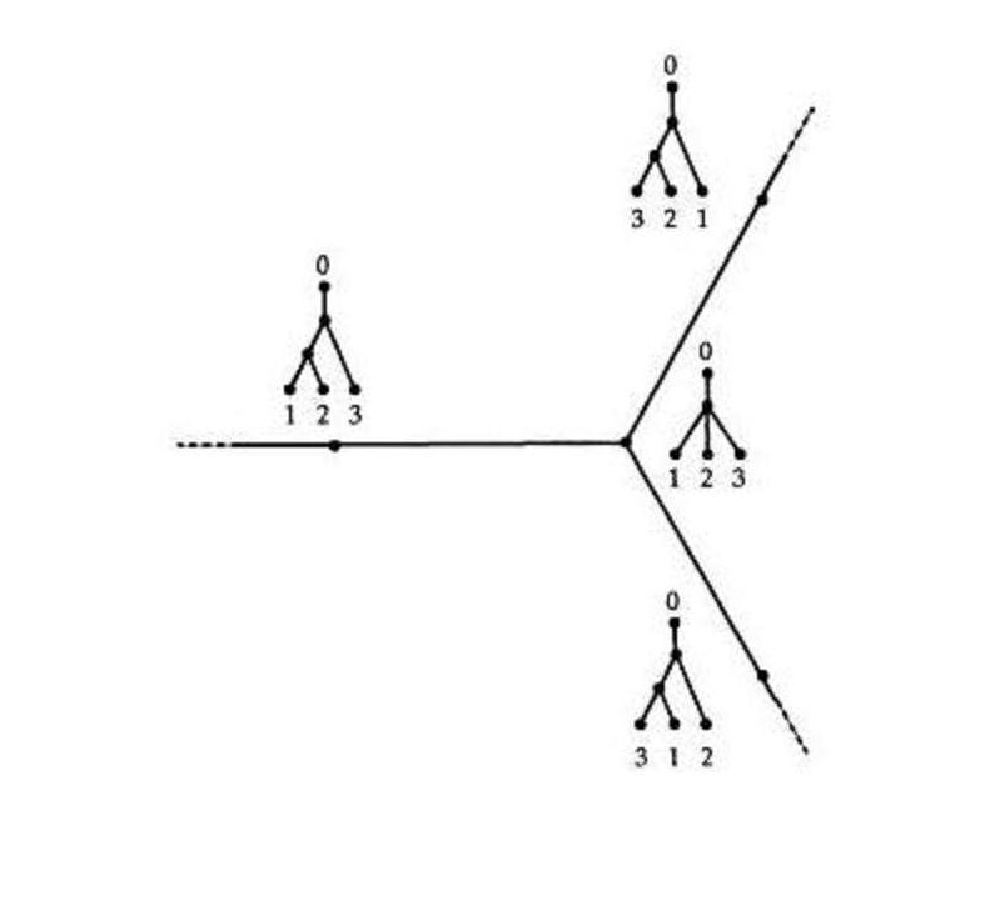}
\includegraphics[scale = 0.20]{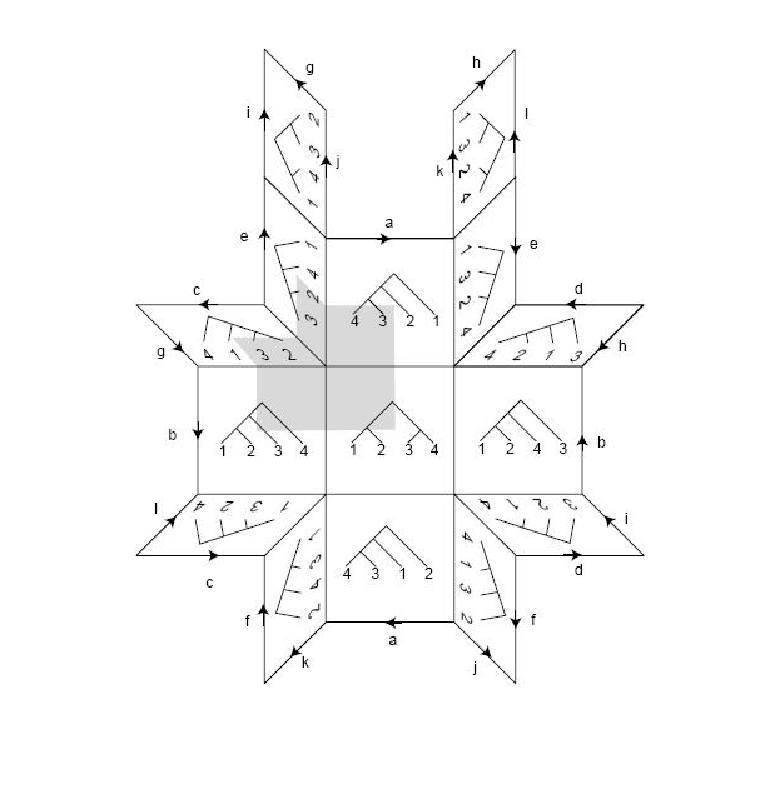}
\includegraphics[scale = 0.20]{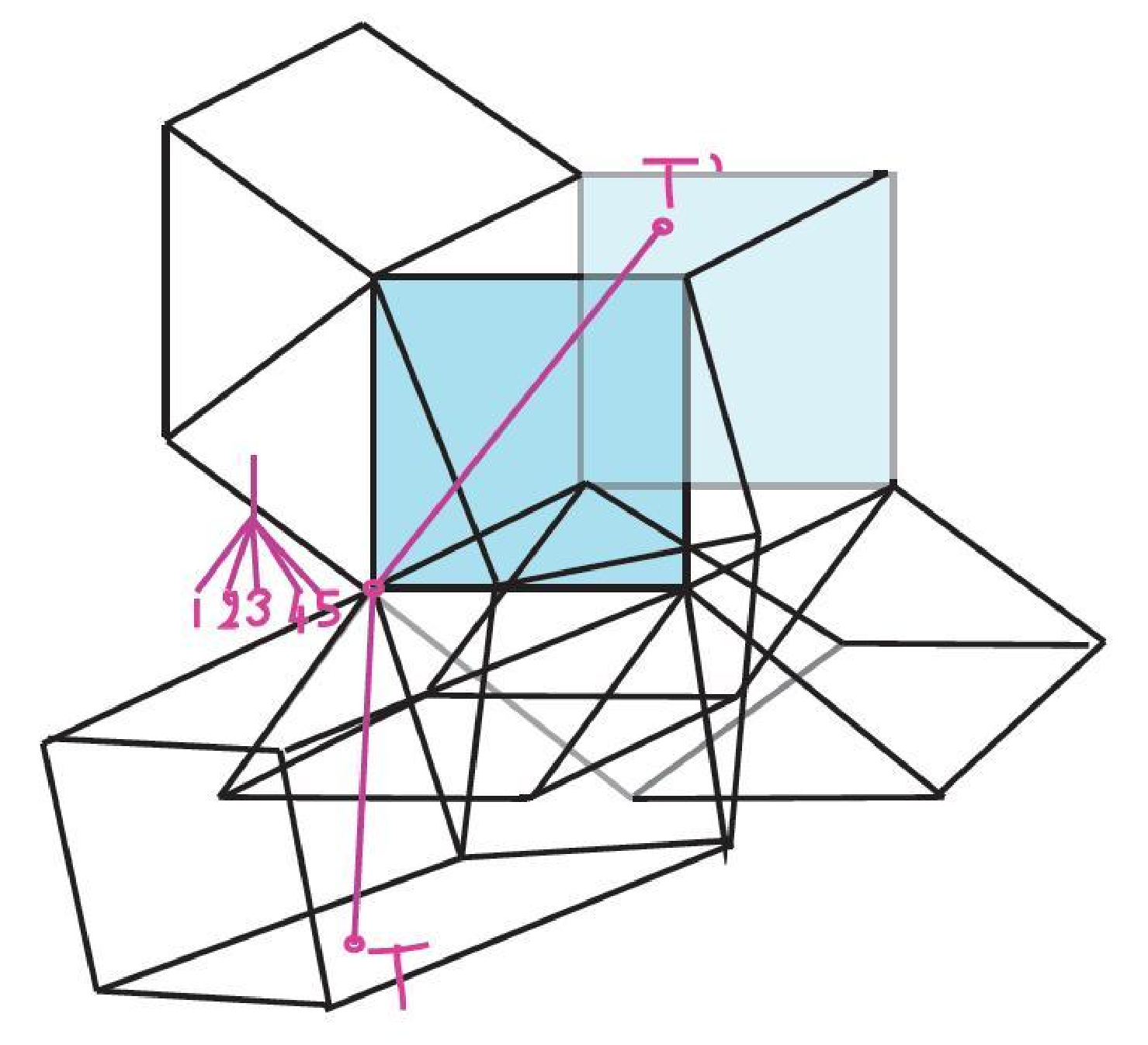}
\caption{Tree spaces $T_3, T_4, T_5$.} \label{f:ts}
\end{figure}

\begin{figure}[H]
\begin{center}
\vspace{-0.45cm}
\includegraphics[scale=0.1]{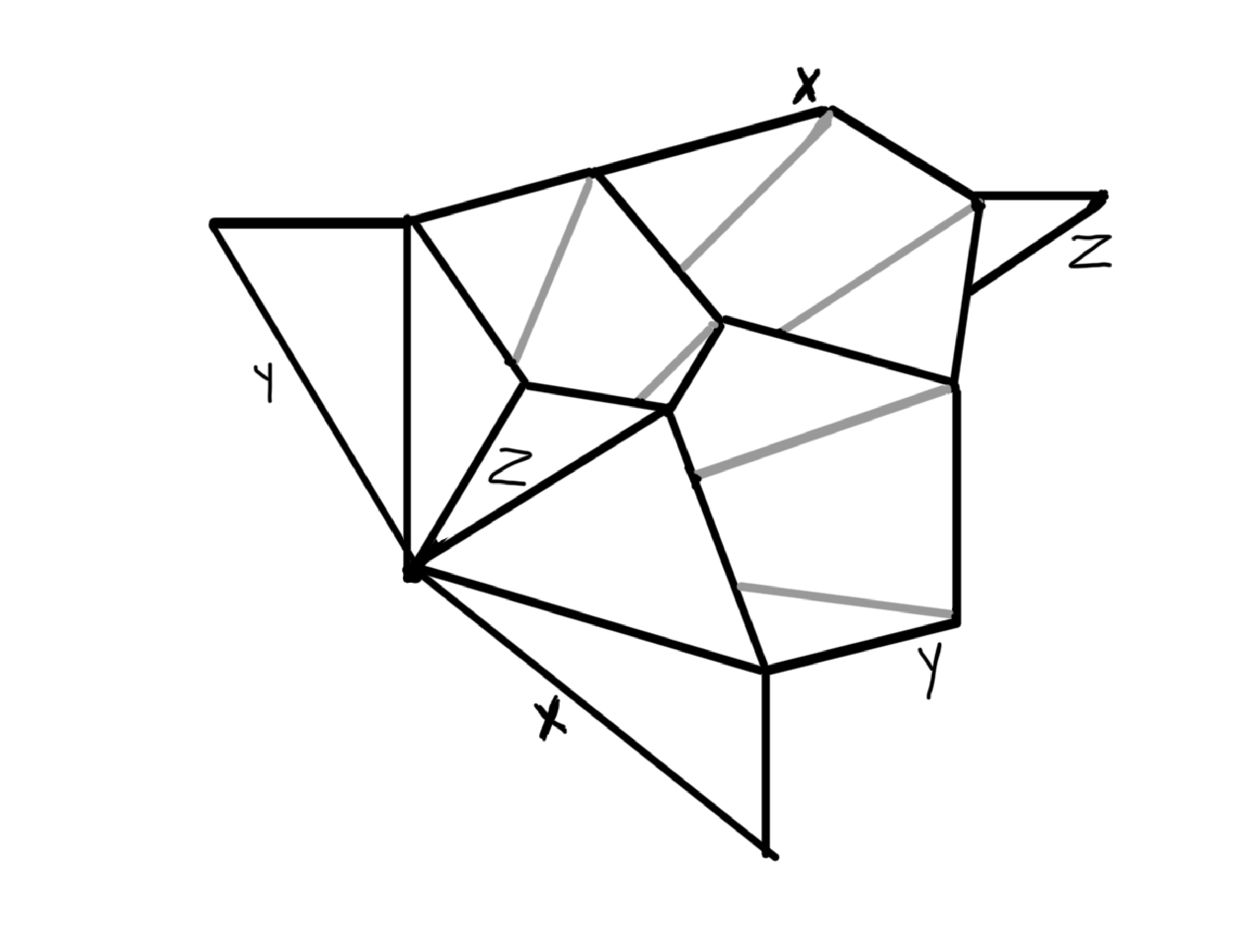}
\caption{A 2D stratified space - $T_4$, space of trees with four leaves} \label{f:t4}
\end{center}
\end{figure}

Additional trees that are more informative or more visually appealing can be found with a quick Google search. If these trees are based on statistical methods, then they necessitate some mathematical background. The more leaves a tree has, the more complicated the space becomes. \\
Billera, Holmes, and Vogtmann \cite{BILLERA:2001} formalized a phylogenetic tree saying a tree with $p$ leaves is an equivalence class based on a certain equivalence, of a DNA-based connected directed graph of species with no loops, having an unobserved {\em root} (common ancestor) and $p$ observed {\em leaves} (current observed species of a certain family of living creatures). In our case, the root is the ancestor language is no longer spoken while the leaves are the modern languages. A tree with $p$ leaves is a simply connected graph with a distinguished vertex, labeled $o$, called the root, and $p$ vertices of degree 1, called leaves, that are labeled from 1 to p. In addition, we assume that all interior edges have positive lengths. An edge of a p-tree is called interior if it is not connected to a leaf.  Now consider a tree T, with interior edges $e_1,\dots, e_r$ of lengths $l_1,\dots, l_r$ respectively. If T is binary, then $r = p-2,$ otherwise $r < n - 2.$ The vector $(l_1,\dots, l_r)^T$ specifies a point in the positive open {\it orthant} $(0, \infty)^r.$  That is to say that a binary $p$-tree has the maximal possible number of interior edges and thus determines the largest possible dimensional orthant; in this case the orthant is $p - 2$-dimensional. The orthant corresponding to each non-binary tree appears as a boundary face of the orthants corresponding to at least three binary trees. In particular, the origin of each orthant corresponds to the (unique) tree with no interior edges, which is known as the {\em star tree}. The space $T_p$ is constructed by taking one $p-2$-dimensional orthant for each of the $(2p - 3)!!$ possible binary trees and gluing them together along their common faces. Note that tree spaces are {\it not} manifolds. Singularities (points where the space does not have a tangent space) are present in the tree space structure. For further detail on phylogenetic trees and the construction of the tree space, see Billera {\em et al.}(2001)\cite{BILLERA:2001}. Phylogenetic trees with $p$ leaves are points on a metric space $T_p$ that has $p-2$ dimensional stratification. In particular, the space of trees with 3 leaves is $T_3 = S_3$, a $3$-spider, which is the union of three line segments with a common end (see Figure \ref{f:ts}, next page). For a probability measure on $S_p,$ if none of the the ``legs" of the $p$-spider has a dominant expected mean distance to the center of the spider, then the Fr\'echet mean is the star tree. This result will be stated more formally in the following section and extended to more general spaces. $T_4$ is a two dimensional stratified space obtained from $15 = (2\times 4 - 3)!!$ 2D quadrants glued according to tree identification rules (see Billera et. al.(2001)\cite{BILLERA:2001}). Interior points of these quadrants are combinatorial binary trees with four leaves, the coordinates of an interior point being given by the two interior edges of a binary tree in one of these combinatorial binary trees. Points on the boundaries of the quadrants, correspond to combinatorial trees with four leaves, which are obtained from a  combinatorial binary tree by shrinking one of the interior edges to zero length.
 Therefore a representation of $T_4$ as a surface with singularities can be obtained from the polyhedral surface given in Figure \ref{f:t4} by identifying the edges labeled with the same letter. While this is a 3D pictorial representation only, in fact, as mentioned in Section 3, given that $2^4 - 4 - 2 = 10,$ $T_4$ is embedded in $\mathbb R^{10}$ having the {\em star tree} at the origin.

\begin{figure}[H]
\begin{center}
\vspace{-0.45cm}
\includegraphics[scale=0.6]{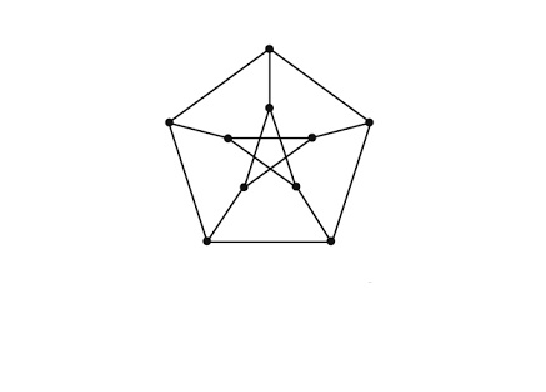}
\caption{Petersen graph} \label{f:Petersen}
\end{center}
\end{figure}

In this representation, the intersection of a small sphere in $\mathbb R^{10}$ centered at the origin with $T_4$ is the so called {\em Petersen graph}. An edge of Petersen graph is the transverse intersection of one quadrant with a sphere, thus there are $15$ edges, and a vertex is the point where one of the coordinate axes pierces the sphere, therefore there are $10$ vertices ( see  figure \ref{f:Petersen}).

For this work, topics like Clustering, Fr\'echet means, and Convergence must be considered to develop some analytical techniques for trees. A brief analysis of two different types of means on tree spaces with 3 leaves will be considered in the context of language ancestry. Finally, new frontiers for analysis of tree spaces will be explored.

\section{Intrinsic Moments on a Metric Space\label{s:2}}

\subsection{General Notions} Fr\`echet's random element in a metric space from 1948 opened new analysis options in the field of statistics. A separable metric space $(\mathcal{M},\rho)$ has some mean with the consistency of the sample mean when the space is compact being shown by Ziezold in 1977. In 2003, the Fr\`echet sample mean and total sample variance on a complete separable metric space were proven to be consistent by Bhattacharya and Patrangenaru \cite{BhPa:2003}. One application of interest here is the random object on a complete space with manifold stratification which will be the phylogenetic tree. Specifically, the intrinsic moments on a manifold $\mathcal{M}$ with probability measure $Q$ result from the Riemannian distance $\rho$ induced by a Riemannian structure $g$ on $\mathcal{M}$.

\subsection{On Stratified Spaces: Open Books}
Hotz, et. al., says the open book is the ``simplest singular topologically stratified space" and defines an open book as the disjoint union of the spine and the interior of the leaves \cite{Hotzetal:2013}. An open book must contain at least 3 leaves ($K\geq 3$). Each leaf contains the non-negative reals; and, when all leaves are ``glued" together at the spine, the leaves comprise just the positive reals with the spine being just the number 0. Hotz, et. al. \cite{Hotzetal:2013} gives the following definition.
\begin{definition}
    The open book $\mathcal{O}$ consists of $K\geq 3$ leaves $L_k$, for $k=1,...,K,$ each of dimension $d+1$ and defined by \begin{equation*}
        L_k=\bar{H}_+\times \{k\}
    \end{equation*} where $\bar{H}_+=\mathbb{R}_{\geq 0}\times S$ with $S=\mathbb{R}^d$. The leaves are joined together along the spine $L_0$ which comprises the equivalence classes in $\bigcup_{k=1}^K(H\times \{k\})$,i.e. $L_0$ can be identified with the hyperplane $H=\{0\}\times S$ or with the space $S=\mathbb{R}^d$.
\end{definition}
Furthermore, the open book $\mathcal{O}$ is the disjoint union of the spine and the interiors $L_k^+=L_k\setminus L_0$ of the leaves so $\mathcal{O}=L_0\cup L_1^+\cup \cdots \cup L_K^+$. On this open book, a metric $d$ (distance) between two points is found by taking the absolute value of the difference between two points if they are on the same or by reflecting ($R$) one of the two points across the hyperplane to a negative value if the two points lie on difference leaves. This reflection is given for $x\in\bar{H}_+$ as $Rx\in\bar{H}_-=\mathbb{R}_{\geq 0}\times \mathbb{R}^d$ Hotz et al give two points $p,q \in \mathcal{O}$ with $p=(x,k)$ and $q=(y,j)$ with $k,j$ denoting the specific leaf of points $x,y$ respectively \cite{Hotzetal:2013}. \begin{equation*}
    d(p,q)= \begin{cases}
        |x-y| & \text{if } k=j, \\
        |x-Ry| & \text{if } k\neq j.
    \end{cases}
\end{equation*}
\subsection{Probability Measures and Moments}
Let $Q=P_X$ be a Borel probability measure on $\mathcal{O}$ such that $d(0,X)$ has bounded expectation and square expectation. For a Borel set $A\subseteq \mathcal{O}$, \begin{equation*}
    Q(A)=w_0Q_0(A\cap L_0)+\sum_{k=1}^K w_kQ_k(A\cap L_k^+)
\end{equation*} where $w$ are weights with $0\geq w_k\geq 1$ and $\sum_{k=1}^Kw_k=1$. \\
To go leaf by leaf, a Folding map $F_k:\mathcal{O}\rightarrow \mathbb{R}^{d+1}$ is used where for $k\in\{1,\ldots,K\}$ and $p\in\mathcal{O}$ \begin{equation*}
    F_kp=\begin{cases}
        x & \text{if } p=(x,k) \in L_k, \\
        Rx & \text{if } p=(x,j) \in L_j \text{ and } j\neq k.
    \end{cases}
\end{equation*}
This allows for the pushforward measure $\tilde{Q}_k=Q\circ F_k^{-1}$ on the $k^\text{th}$ leaf to be \begin{equation*}
    \tilde{Q}_k(A)=w_kQ_k(A\cap \bar{H}_+)+w_0Q_0(A\cap S)+\sum_{j\geq 1,j\neq k} w_jQ_j(A\cap H_-).
\end{equation*}
With the measures given, moments can be found for each leaf. Hotz et al defines the first moment on a leaf \cite{Hotzetal:2013}. \begin{definition}
    Let $x^{(0)},x^{(1)},\ldots,x^{(d)}$ be the coordinate functions on $\mathbb{R}^{d+1}$. The first moment of the measure Q on the $k^\text{th}$ leaf $L_k$ is the real number \begin{equation*}
        m_k=\int_{\mathbb{R}^{d+1}}x^{(0)}d\tilde{Q}_k(x)=\int_{\mathcal{O}}(\pi_0F_kp)dQ(p),
    \end{equation*} where $\pi_0:\mathbb{R}^{d+1}\rightarrow \mathbb{R}$ is the orthogonal projection with kernel $H=\{0\}\times \mathbb{R}^d$.
\end{definition}
This first moment can only be nonnegative on at most one leaf as proven by Hotz et al. (2013)\cite{Hotzetal:2013}, where the following definition is given.
\begin{definition}
    Under integrability and nondegeneracy, the mean of the measure $Q$ is either \begin{enumerate}
        \item nonsticky if $m_k>0$ for some (unique) $k\in\{1,\ldots,K\}$, or
        \item partly sticky if $m_k=0$ for some (unique) $k\in\{1,\ldots,K\}$, or
        \item sticky if $m_j<0$ for all indices $j\in\{1,\ldots,K\}$.
    \end{enumerate}
\end{definition}

When the dimensionality of the data is one, the open book becomes a spider. When $k=3$, it is the simplest open book called a 3-spider.

\section{Sample Mean on $\mathcal{T}_3$ \label{s:3}}
\begin{figure}[h]
\begin{center}
\includegraphics[scale = 0.4]{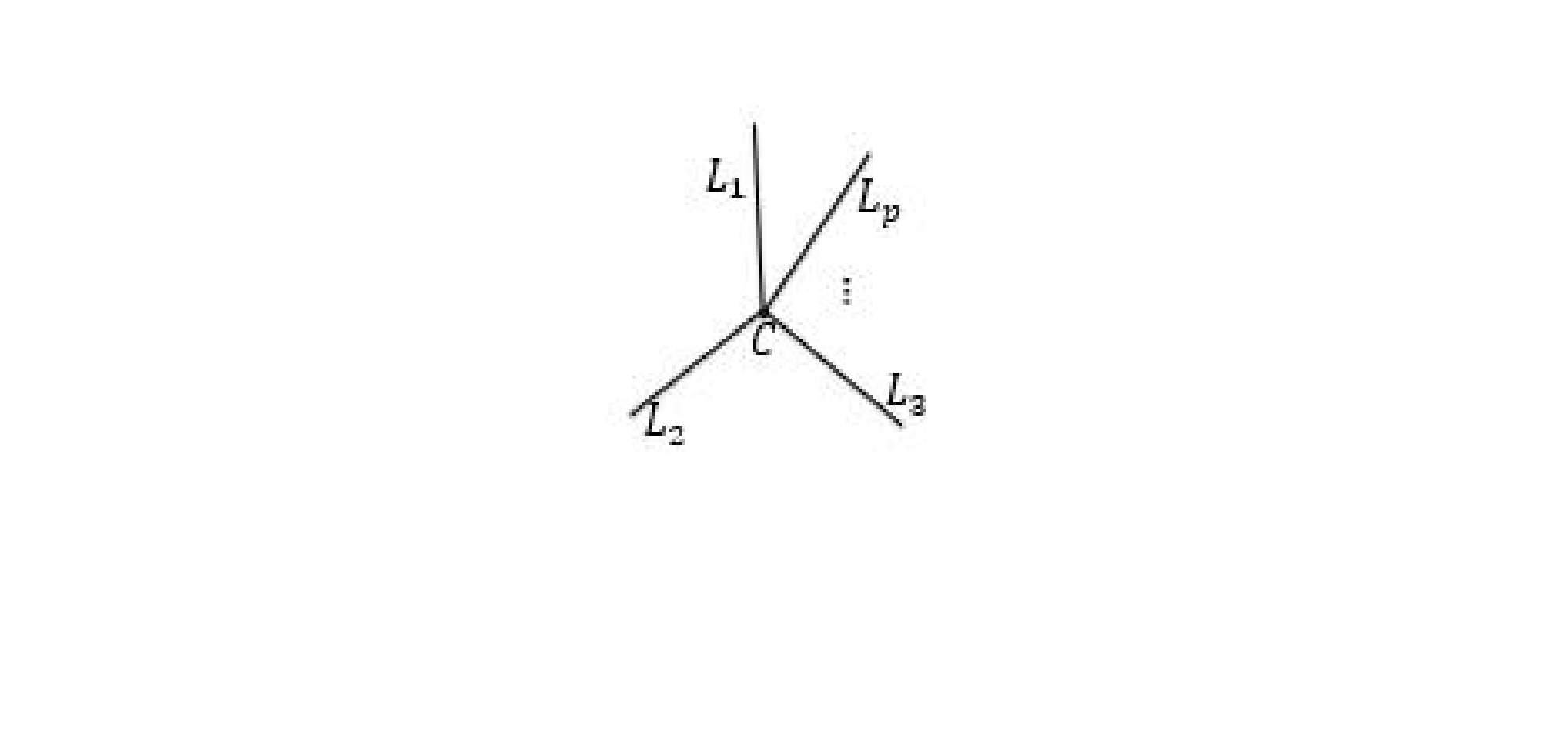}
\end{center}
\caption{Spider $S_p$}%
\label{fig:spider}%
\end{figure}

Results in this section are presenting  an elementary proof of the {\em Sticky CLT} in Hotz et al(2013)\cite{HHLMMMNOPS:2013}. To define a Spider, consider an arbitrary nonempty set $K$ of size at least $3$ and, for each of its elements, $i$, define the ray(leg) $L_i=\{(i,x):x\in [0,\infty)\}$. The Spider is formed by identifying the origins $(i,0),$ for all $i \in K,$ which corresponds to joining the rays together at a common center $C$ (see figure \ref{fig:spider}).

\[
S_K=\{(i,x):i\in K,x\in [0,\infty )\}
\]
where $(i,0),\dots,(j,0)$ for all $i,j \in K$, the equivalence class of all points of the form $(i,0)$, we denote by $C$, named {\em center of the spider}.

Assume $X_i,i=1,\dots,n$ are i.i.d. random objects on a spider $S_K$, having legs $L_i,i=1,\dots,K$ and center $C$. Further, assume the intrinsic mean $\mu_I$ exists and the intrinsic variance is finite. The sample mean of each leaf must be found by finding what Hotz et al calls the $k^th$ folded average \begin{equation*}
    \eta_{k,N}=\frac{1}{N}\sum_{n=1}^N F_kp_n
\end{equation*}
for $N$ points $\{p_n\}_{n=1}^N\subset \mathcal{O}$. Following from the theory presented earlier, at most one of these folded averages can be nonnegative. \\
Any probability measure $Q$ on $S_K$ then decomposes uniquely as a weighted sum of probability measures $Q_K$ on the legs $L_K$ and an atom $Q_0$ at $C$. More precisely, there are nonnegative real numbers $\{w_k\}_{k=0}^p$ (probability weights) summing to 1 such that, for any Borel set $A\subseteq S_p$ , the measure $Q$ takes the value
\[
Q(A)=w_0Q_0(A\cap C)+\sum_{k=1}^p w_k Q_k(A\cap L_k).
\]

Assume $w_0=0$ and $x\in L_a$, the Fr\'echet function $F_X(x) = F(x) = E(\rho^2(X,x)).$ If there exists an unique minimizer for the Fr\'echet function $F(x)$, the minimizer is called intrinsic mean $\mu_I$ (based on the intrinsic distance $\rho$). If on each leg, the distance is Euclidean, $F(x)$ is defined as follows,
\begin{align*}
 F(x) &= \sum_{i=1,i\neq a}^K \int_0^{\infty} (x+u)^2w_iQ_i(du) + \int_0^{\infty}(x-u)^2w_aQ_a(du) \\
 &= x^2\sum_{i=1}^K \int_0^{\infty}w_iQ_i(du) +2x[\sum_{i=1,i\neq a}^K  \int_0^{\infty} uw_iQ_i(du) -  \int_0^{\infty} uw_aQ_a(du)]+ \sum_{i=1}^K  \int_0^{\infty}u^2w_iQ_i(du)\\
 &=x^2+2[\sum_{i=1,i\neq a}^K v_i - v_a]x +const.
\end{align*}
where $v_i= \int_0^{\infty}uw_iQ_i(du)$.

The minimizer of the quadratic form is the first sample moment denoted $x^*_a=v_a-\sum_{i\neq a}v_i$, where $x_a \in L_a =\{(a,u):u\in [0,\infty)\}$. Thus, we have three situations:(i) $x^*_a>0$ for some unique $a\in \{1,...,K\}$, (ii) $x^*_a=0$ for some unique $a\in \{1,...,K\}$, and (iii) $x^*_a<0$ $\forall a$. In case (i), we have $\mu_I$ well defined on $L_a$, then the classical C.L.T is applied. In case (ii), we can fold the other legs into that half line opposite to $L_a$ and then apply C.L.T. Since the negative part is undefined, so the result goes to a positive truncated normal distribution. In case (iii), for any $a\in \{1,\dots,K\}$, we have $\mu_I=C$, which shows that intrinsic mean $\mu_I$ sticks to the center $C$.

\begin{enumerate}
  \item $x^*_a>0$ for some (unique) $a\in \{1,\dots,K\}$, then $\mu_I \in L_a$ and  due to the consistency of the intrinsic sample mean, for $n$ large enough $\bar{X}_{I,n} \in L_a$ and $\sqrt{n}(\bar{X}_{I,n}-\mu_I)$ has asymptotically a normal distribution.
  \item $x^*_a=0$ for some (unique) $a\in \{1,\dots,K\}$, then after folding the legs $L_i,i\neq a$, into one half line opposite to $L_a$, $\sqrt{n}(\bar{X_n}-\mu_I)$ has asymptotically a positive truncated normal distribution.
  \item $x^*_a<0$ for all $a\in \{1,\dots,K\}$, then $\mu_I=C$ and there is $n_0$ s.t. $\forall n \geq n_0$, then $\bar{X}_{I,n}=C$ a.s.
\end{enumerate}

\section{Applications to Language Ancestry\label{s:4}}
\subsection{Data Collection}
\subsubsection{Words}
Of the varieties of cognate lists for language ancestry, one of the original lists developed by Morris Swadesh contains 207 words \cite{Swadesh:1952}\cite{Swadesh:1955}\cite{Swadesh:1971}\cite{wiki:2023}. These words range from familial relations and numbers to action verbs and anatomy. Each language has a representation of these words. The lists were accessed through the Wikipedia Module ``Swadesh modules" which contains the 207 words in hundreds of languages. The languages chosen for analysis are English, German, French, Italian, Spanish, and Irish (Gaelic). Words were input into a data frame in R for analysis. To imagine two populations, words were randomly sampled into groups of 102 and 105. Words were then randomly sampled 3 at a time to create 34 samples from population 1 and 35 from population 2. Three analyses will be run involving English, German, and French first then French, Italian, and Spanish second, and German, French, and Irish last.
\subsubsection{Calculate Distances}
Per Johnson and Wichern's example \cite{JohnsonWichern:2019}, the first letter of each word is used to check for similarity. If the letter is the same, then a distance of zero is assigned. If the letter is different, then a distance of one is assigned. This is done for each of the 207 words between each pair of languages to be analyzed: English and German, English and French, German and French, etc. Sum the distances within each of the 69 samples for two languages at a time. The minimum possible distance is 0 when all 3 languages begin with the same letter for all 3 words and the maximum possible distance is 3 when all 3 languages begin with different letters for all 3 words. This produces a $3\times 3$ distance matrix. \\
Some examples of the distance calculations are given next.

\begin{example}\index{d1}\label{ex:d1} This is an example where all three languages are equidistant in the sample.
\end{example}

\includegraphics[scale=.25]{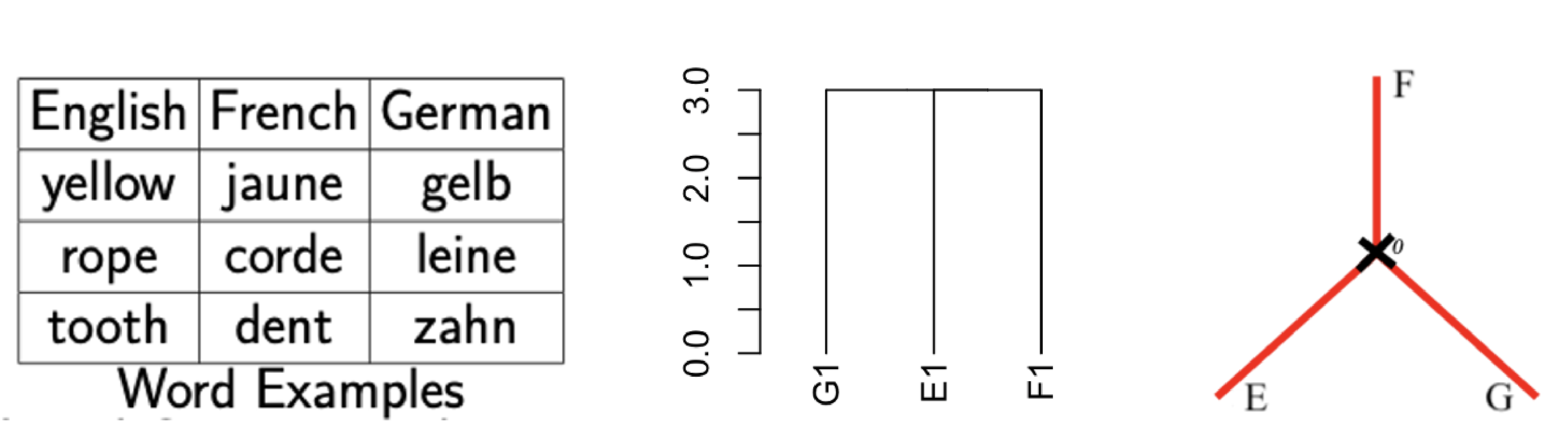}

%%\begin{tabular}{cc|c|c} \hline
%%    & English & French & German \\ \hline
%%    & yellow &  jaune  & gelb\\ \hline
%%    & rope &  corde  &  leine\\ \hline
%%    & tooth &  dent  & zahn \\ \hline
%%    \\
%%    \\
%%    \\
%%\end{tabular} \quad \quad
%%\includegraphics[scale=0.7]{dendro_equi.eps}
%%\includegraphics[scale=0.5]{origin.eps} \\

In this example, the distance between English and French is 3, between English and German is 3, and between French and German is 3.
\begin{example}\index{d2}\label{ex:d2}
This is an example where one language is furthest away and equidistant from the other two.
\end{example}

\includegraphics[scale=0.25]{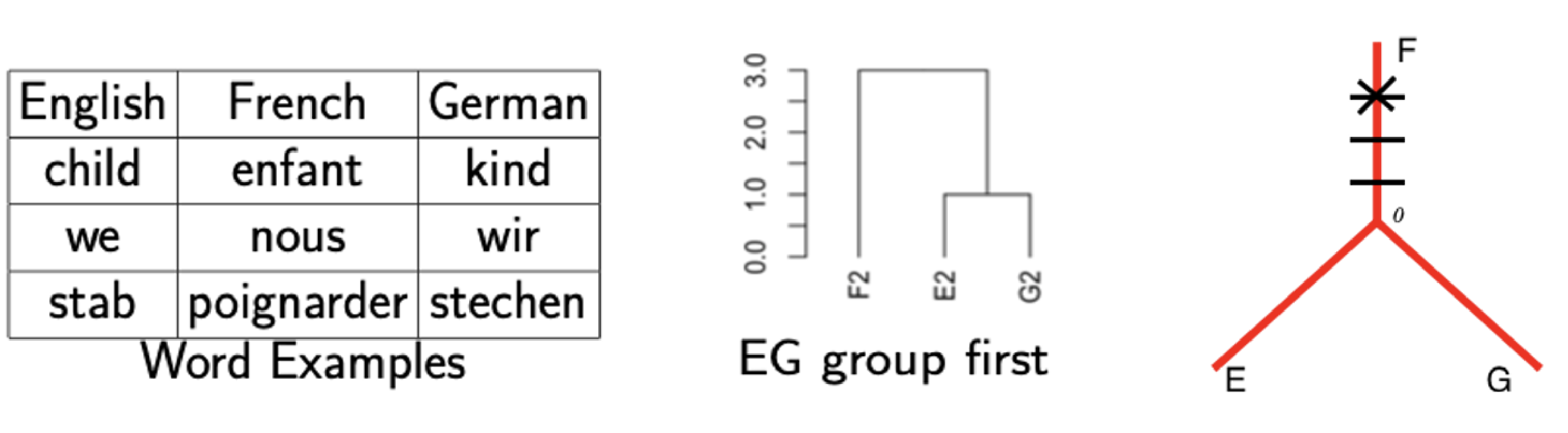}

%%\begin{tabular}{cc|c|c} \hline
%%    & English & French & German \\ \hline
%%    & child &  enfant  & kind\\ \hline
%%    & we &  nous  &  wir\\ \hline
%%    & stab &  poignarder  & stechen \\ \hline
%%    \\
%%    \\
%%    \\
%%\end{tabular}  \quad \quad
%%\includegraphics[scale=0.7]{dendro_f3.eps}
%%\includegraphics[scale=0.5]{f_3dist.eps} \\

In this example, the distance between English and French is 3, between English and German is 1, and between French and German is 3.
\begin{example}\index{d3}\label{ex:d3}
This is an example where one language is furthest away, but is closer to one language than it is to the other.
\end{example}

\includegraphics[scale=0.23]{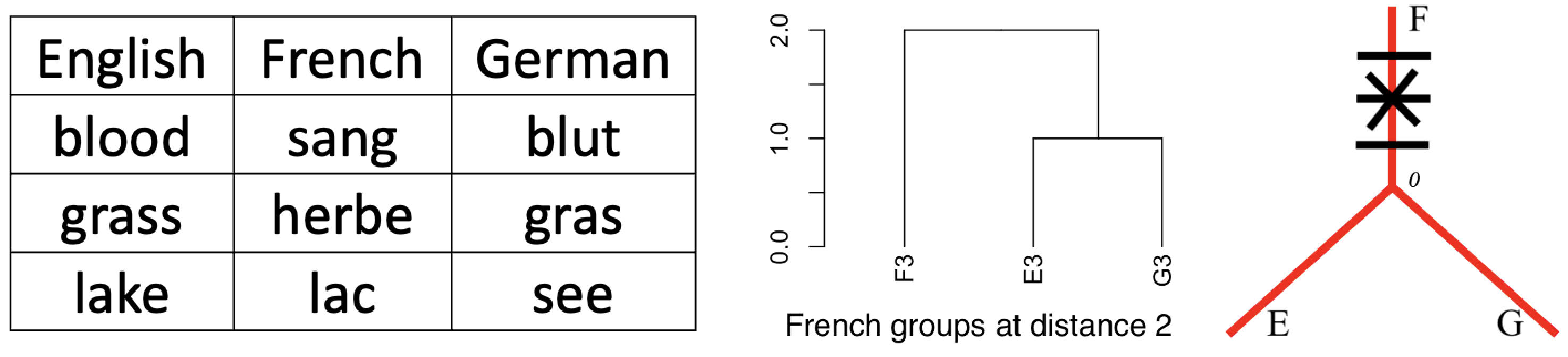}
%%\begin{tabular}{cc|c|c} \hline
%%    & English & French & German \\ \hline
%%    & blood &  sang  & blut\\ \hline
%%    & grass &  herbe  &  gras\\ \hline
%%    & lake &  lac  & see \\ \hline
%%    \\
%%    \\
%%    \\
%%\end{tabular} \quad \quad
%%\includegraphics[scale=0.35]{dendro_f2.eps}
%%\includegraphics[scale=0.25]{spider_f2.eps} \\

In this example, the distance between English and French is 2, between English and German is 1, and between French and German is 3.
\begin{example}\index{d4}\label{ex:d4}
This is an example where two groups of languages tie for the minimum distance. No graphs are shown since more than one graph could be created depending on the random choice between clustering English and French first or English and German first.
\end{example}
\begin{tabular}{c|c|c} \hline
    English & French & German \\ \hline
    road &  route  & stra{\ss}e\\ \hline
    freeze &  geler  &  frieren\\ \hline
    because &  parce que  & weil \\ \hline
\end{tabular} \\

In this example, the distance between English and French is 2, between English and German is 2, and between French and German is 3.
\subsubsection{Points on a 3-spider}
The distances are used to cluster as discussed in the Preliminaries for single linkage and then given as points on a 3-spider. Here, the three individuals are three languages with the following four possible ways to cluster and assign points on leaves. \begin{enumerate}
    \item All languages are equidistant so only one cluster is needed. In the context of a 3-spider, the point is at the origin. See \ref{ex:d1}.
    \item One language is furthest away and is equidistant from the other two. For a 3-spider, the point is on that language's leaf with a distance equal to the distance to the other languages. See \ref{ex:d2}. The point is at 3 on the French leaf.
    \item One language is furthest away, but is closer to one language than the other. For a 3-spider, the point is on that language's leaf with a distance equal to the minimum distance from that language to the other two. See \ref{ex:d3}. The point is 2 on the French leaf.
    \item Two groups of languages tie for the minimum distance. For example, Language 1 is distance $c_1$ from Language 2 and Language 2 is also distance $c_1$ from Language 3, but Language 1 is a distance greater than $c_1$ (say $c_2$) from Language 3. In this case, \textit{randomly} decide whether to cluster 1 with 2 or 2 with 3. For a 3-spider, then use the $c_2$ distance as the point on the Language's leaf that was not clustered originally (the point would be at $c_2$ on leaf 1 or leaf 3). See \ref{ex:d4}. The point is 2 and was randomly assigned to the French leaf.
\end{enumerate}
The distances are then points on a 3-spider where a point is on a particular leaf $L_i$ where $i=E,G,F$ or $i=F,I,S$ if that language is the furthest (added to the cluster last).
\subsection{Non-sticky Sample Mean}
English, French, and German are the first three chosen languages. English is classified as a Germanic language while French is classified as a Romance (Italic or Latin) language. As discussed in the previous sections, only one leaf may have a positive mean (although none are guaranteed). In this case, one language seems to be more different than the other two.\\

\begin{tabular}{|c|c|c|} \hline
    Language & Mean Distance 1 & Mean Distance 2 \\ \hline
    English &  -2.00  & -2.41\\ \hline
    German &  -1.66  &  -1.41\\ \hline
    French &  1.49  & 1.41 \\ \hline
\end{tabular} \\

Since French has a positive mean distance, it is considered the sample mean (barycenter) and is used to calculate the measures of variability. This is the non-sticky case for the mean. The variance of the first sample is 4.14 and the second is 4.61. Using the Non-sticky Central Limit Theorem, traditional asymptotic inference can be performed. The pooled sample variance is 4.43. These statistics yield the two sample t-statistic value of 0.16 which is far from significant. This is to be expected since the samples had no necessary distinction; they were randomly assigned. If two groups of words were different for some reason, say medical terms versus agricultural terms, then perhaps some significant difference may be found. \\

\subsection{Sticky Sample Mean}
French, Spanish, and Italian are 3 popular Romance languages sharing many commonalities. Latin was the language of the ancient boot peninsula and used prominently by the Roman Catholic church for centuries. However, with no speakers today, Latin is considered a dead language. In this case, we may expect many observations at the origin and no starkly different language.\\

\begin{tabular}{|c|c|c|} \hline
    Language &  Mean Distance 1  & Mean Distance 2 \\ \hline
    French  & -0.51 &  -0.76\\ \hline
    Spanish  & -0.74 &   -0.35\\ \hline
    Italian  & -0.11 &  -0.05 \\ \hline
\end{tabular} \\

Since none of the means are positive, the sample mean is considered to be sticky, and the barycenter is at 0. From that mean of zero, the variances of each sample are calculated to be 2.88 and 2.30 respectively.  \\

\subsection{Another Sticky Sample Mean}
So far, two cases have been considered where one language has been from a different ancestor and where all three languages are from the same ancestor. This case will consider all three languages from different ancestors. German, French, and Irish (Gaelic, from the Celtic ancestor) words from the Swadesh 207 list were used from the earlier procedures to find the following mean distances.

\begin{tabular}{|c|c|c|} \hline
    Language & Mean Distance 1 & Mean Distance 2 \\ \hline
    German &  -0.74  &  -0.03\\ \hline
    French &  -0.46  & -0.97 \\ \hline
    Irish & -0.40 & -0.85\\ \hline
\end{tabular} \\

In both samples, all 3 means are negative values indicating a sticky sample mean to be placed at the origin. From that mean of zero, the sample variances are calculated to be 4.88 and 5.45, respectively. Unlike the previous case of a sticky sample mean where the languages were considered to be of the same ancestor, here, the languages are believed to all be of different ancestors.
\section{Conclusions\label{s:5}}
The clustering of language may be done through the use of the 3-spider. In the three applications shown, a non-sticky mean was observed when one language was from a different family while a sticky mean was observed when all languages have the same ancestry or all languages has different ancestry. English and German are from the Germanic family while French, Italian, and Spanish are from the Latin (Romance, Italic) family while German, French, and Irish are from different families. For these examples, the samples representing two populations were simply randomly assigned, but if two groups of words were different for some reason, say medical terms versus agricultural terms, then perhaps some significant difference may be found in the ancestry as a different leaf may dominate in each case. Other applications may be more apparent to a studied linguist. While these results may not apply to all languages, it provides a start. Furthermore, inference may be done using the Central Limit Theorems for each case. \\
The distance calculated between each language was also exceptionally simple. An edit distance for full words may provide additional insights. While the Swadesh list of words has been the standard for linguistics, phrases or sentences introducing structure could be of interest for investigation. The languages in this study all used the Latin alphabet which is extremely common in Western civilization, but it is not the only alphabet as Cyrillic, Greek, and Arabic among others could be considered. Paradoxically, although while here compare Indo-European languages, our primary tool, the Latin alphabet sources from the Proto-Sinaitic script developed from the popular non Indo-European language of Semitic-speaking workers and slaves in the antique Egypt, as they were unskilled to handle the hieroglyphic system used in the Egyptian language, that required a large number of pictograms, as opposed to the semantic values, of the own Canaanite language of slaves. This Semitic alphabet, ancestor of multiple writing systems across the Middle East, imposed itself having only about half as many letters as the Vedic Sanskrit alphabet, who appeared around the same period. The written language only was considered regardless of pronunciation which would need the International Phonetic Alphabet (IPA) instead of the Latin Script. With the addition of more than 3 languages, additional types of phylogenetic trees beyond the 3-Spider would need to be considered. This grows the number of possible trees quickly following from the work by Billera and Holmes necessitating additional computational power and theory.

\section{Acknowledgements}
The authors acknowledge support by the National Science Foundation under Grant DMS:2311059.

\section{R Code with Data}
Please access the R file at the following github repository: \\
\url{https://github.com/GarettO9/lang_tree/tree/main}.

The code contains the data and commands with a set seed for random selection to reproduce the results.

\bibliographystyle{plain}
\bibliography{Lang_paper_04_09_2024}

\end{document}